\documentclass[manuscript]{aastex}
\usepackage{multirow}
\usepackage{amsmath}
\usepackage{amssymb}
\usepackage{epsfig}
\usepackage{graphicx}
\usepackage{subfigure}
\usepackage{epstopdf}
\usepackage{longtable}
\usepackage{lineno}

\shorttitle{Solar-cycle Variability of Solar Wind Turbulence}
\shortauthors{Fa \& He}

\begin{document}
\title{Solar-cycle Variability of Composite Geometry in the Solar Wind Turbulence \\}
\author{Zhan Fa\altaffilmark{1,2} and H.-Q. He\altaffilmark{1,2}}

\altaffiltext{1}{Key Laboratory of Planetary Science and Frontier
Technology, Institute of Geology and Geophysics, Chinese Academy of
Sciences, Beijing 100029, China; hqhe@mail.iggcas.ac.cn}

\altaffiltext{2}{College of Earth and Planetary Sciences, University
of Chinese Academy of Sciences, Beijing 100049, China}

\begin{abstract}
The composite geometry and spectral anisotropy of the solar wind
turbulence are very important topics in the investigations of solar
wind. In this work, we use the magnetic field and plasma data from
Wind spacecraft measured during 1995 January to 2023 December, which
covers more than two solar cycles, to systematically investigate
these subjects in the context of solar-cycle variability. The
so-called spectrum ratio test and spectrum anisotropy test are
employed to determine the three-dimensional (3D) geometry of the
solar wind turbulence. Both the tests reveal that the solar wind
turbulence is dominated by the two-dimensional (2D) component
($\sim80\%$ by turbulence energy). More interestingly, we find that
the fraction of slab turbulence increases with the rising sunspot
number, and the correlation coefficient between the slab fraction
and the sunspot number is 0.61 (ratio test result) or 0.65
(anisotropy test result). This phenomenon suggests that the
increasing solar activity (signified by sunspot number) causes
increasing slab component in the solar wind turbulence. The
relationship between spectral anisotropy and solar activity is
discussed and explained. The enhancement of slab fraction is
associated with the intensified interplanetary magnetic field
magnitude and the increased Alfv{\'e}n speed during the rise phases
of the solar cycles. Our findings will be very helpful for achieving
a better understanding of the 3D composite geometry and spectral
anisotropy of the solar wind turbulence, and especially of their
solar-cycle variability.
\end{abstract}


\section{Introduction} \label{sec:intro}
Interplanetary space, filled with magnetized plasmas, can serve as a
unique laboratory for investigating diverse space plasma phenomena.
As the primary stellar wind directly detectable, the solar wind
provides us an invaluable opportunity to study magnetohydrodynamics
(MHD) turbulence \citep{Tu1995,Bruno2013}. The solar wind turbulence
is related to several crucial topics regarding the physical
processes in the interplanetary space, such as energy transfer,
momentum exchange, and dynamics of energetic particles
\citep{Jokipii1970,Goldstein1995,Velli2003,He2015,He2021}. Recently,
the anisotropy of solar wind turbulence has been intensely
investigated in the community of space physics and solar physics.

A prominent feature of solar wind turbulence is its anisotropy,
which arises from the presence of a mean magnetic field that
introduces a preferred direction within the plasmas
\citep{Horbury2008,Wicks2010}. The anisotropy is related to the
spatial distribution of turbulent fluctuations and their spectral
properties. As a pivotal aspect of solar wind turbulence, spectral
anisotropy has drawn considerable attention. The spectral anisotropy
is crucial as it governs the cascade of energy from large to small
scales and affects the dynamics of wave-particle interactions
\citep{Goldreich1995,Chen2016}. There are three primary models for
describing the geometry of solar wind turbulence: the slab model,
the two-dimensional (2D) model, and the slab/2D two-component
composite model. The so-called slab model posits that the
fluctuations propagate strictly parallel to the mean magnetic field
\citep{Parker1979,Zhou1990}. The 2D model assumes that wavevectors
are oriented perpendicular to the mean field, capturing the
predominant transverse properties of observed fluctuations
\citep{Shebalin1983,Oughton1994}. However, empirical observations
have consistently demonstrated that neither the slab model nor the
2D model alone sufficiently captures the complexity of solar wind
turbulence \citep{Matthaeus1990,Bieber1996}.

\citet{Matthaeus1990} identified significant discrepancies between
theoretical predictions based on pure slab turbulence and
observations of cosmic-ray propagation. These discrepancies
necessitate incorporating 2D component to reconcile the theoretical
models with observational results. Consequently, the two-component
composite models combining slab and 2D components were proposed and
subsequently supported by observational investigations
\citep{Bieber1996,Dasso2005}. Further insights into the composite
property of solar wind turbulence were provided by
\citet{Matthaeus1995}, who recognized the combined influences of
slab and 2D components. \citet{Bieber1996} analyzed the data from
Helios 1 and Helios 2 spacecraft, and suggested that the solar wind
turbulence for a frequency band in the inertial range is
2D-component dominated ($\sim84\%$ of the total turbulence energy).
\citet{Oughton2015} and \citet{Oughton2021} also reported that the
2D component is predominant in the inertial range, with energy
fraction ranging from $50\%$ to $100\%$.

Utilizing axisymmetric models, \citet{Bieber1996} established two
methods, i.e., spectrum ratio test and spectrum anisotropy test, to
infer the relative significance of 2D component in the solar wind
turbulence during 29 solar proton events. The spectrum ratio test,
based on the ratio of perpendicular to parallel spectra, suggested a
composite structure with $26\%$ slab component and $74\%$ 2D
component for inertial range in solar wind turbulence. The spectrum
anisotropy test evaluated the dependence of spectra on the angle
between the mean magnetic field and the solar wind velocity. As the
angle increases, the parallel spectrum component decreases, while
the perpendicular component increases. The anisotropy test suggested
a composite structure with $5\%$ slab component and $95\%$ 2D
component. Similar results were obtained by \citet{Zank1992} in Mach
number scaling experiments in the frame of incompressible theory.
They found that approximately $80\%$ of turbulence energy is
distributed in the 2D component and $20\%$ in the slab component.

The anisotropy of solar wind turbulence is influenced by solar wind
conditions. \citet{Dasso2005} found that the parallel turbulence
component is significant in fast solar winds, while the
quasi-perpendicular (2D) turbulence component is prominent in slow
solar winds. \citet{Narita2018} reviewed the wavevector distribution
of turbulence energy in different directions, and implied that
spectral anisotropy may be influenced by the instability of the mean
magnetic fields and plasmas. \citet{Osman2007} obtained spatial
autocorrelation functions using two-point measurements of magnetic
fields from Cluster spacecraft, and gave the value $1.79\pm0.36$ of
the ratio of correlation scales along the mean field direction to
those along the field-perpendicular directions. Recently,
\citet{Zhou2020} investigated the solar-cycle variability of
correlation scale, Taylor scale, and effective magnetic Reynolds
number in the solar wind turbulence. They found that the correlation
coefficient between the Taylor scale and the sunspot number is 0.92,
and the correlation coefficient between the effective magnetic
Reynolds number and the sunspot number is -0.82. \citet{Zhou2021}
explored the solar-cycle variability of the anisotropy of solar wind
turbulence. Their results revealed that the correlation coefficient
between the anisotropy ratios of the Taylor scales and the sunspot
number is 0.65, and the correlation coefficient between the
anisotropy ratios of the effective magnetic Reynolds number and the
sunspot number is -0.75.

In this work, we use the data measured by Wind spacecraft during the
long time period 1995 January to 2023 December to investigate the
long-term (solar-cycle) variability of the composite geometry and
spectral anisotropy of the solar wind turbulence. This paper is
arranged as follows. In Section \ref{sec:Models for Anisotropic
Spectra}, we provide a brief description of the theoretical bases
for the two analysis methods of observational data, i.e., the
so-called spectrum ratio test and spectrum anisotropy test. Section
\ref{sec:Data and Methods} describes the data selection criteria,
processing techniques, and utilization of the spectral analysis
methods. The results are presented and discussed in Section
\ref{sec:Results and Discussion}, where we examine the influence of
solar activity on solar wind turbulence anisotropy. A summary of our
findings will be provided in Section \ref{sec:Summary}.

\section{Models for Anisotropic Spectra} \label{sec:Models for Anisotropic Spectra}
We employ a single-spacecraft technique to determine the 3D
anisotropic geometry of solar wind turbulence. This method has been
used in a series of works
\citep[e.g.,][]{Bieber1996,Horbury2008,Forman2011}. We establish a
right-handed orthogonal coordinate system for investigating the
axisymmetric turbulence concerning the local mean magnetic field
$\mathbf{B}_0$: the $z$-axis is aligned with $\mathbf{B}_0$, the
$x$-axis is set to be coplanar with both $\mathbf{B}_0$ and the
solar wind flow velocity $\mathbf{V}$, and the $y$-axis completes
the right-handed orthogonal coordinate frame. The angle $\psi$
denotes the acute angle between $\mathbf{B}_0$ and $\mathbf{V}$.

For spatially homogeneous turbulence, the power spectral density
tensor $P_{ij}(\mathbf{k})$ can be expressed using the two-point
correlation tensor $R_{ij}(\mathbf{r})$ as
\begin{equation} \label{eq1}
P_{ij}(\mathbf{k})=\frac{1}{(2\pi)^3}\int
R_{ij}(\mathbf{r})e^{-i\mathbf{k}\cdot\mathbf{r}}d^3\mathbf{r}.
\end{equation}
Here, $\mathbf{k}$ and $\mathbf{r}$ denote the wavevector and the
spatial separation vector, respectively. Within the plasma
framework, the spacecraft velocities are significantly lower than
the solar wind speeds $V$, and hence, the Taylor hypothesis
\citep{Taylor1938} is usually applied to relate the local spatial
and temporal correlation tensors. Using the material derivative
$\mathrm{D}/\mathrm{D}t={\partial}/{\partial t}+V\cdot\nabla$, we
can obtain the relation
\begin{equation}\label{eq2}
R_{ij}^{(t)}(\tau)=R_{ij}^{(x)}\left(-V\sin\psi\tau,0,-V\cos\psi\tau\right).
\end{equation}
Here, $R_{ij}^{(t)}(\tau)$ is the temporal correlation tensor at
time lag $\tau$, and $R_{ij}^{(x)}(\mathbf{r})$ is the spatial
correlation tensor.

Considering the magnetic fluctuations
$\mathbf{b}=\mathbf{B}-\mathbf{B}_0$ with
$\langle\mathbf{b}\rangle=0$, where $\langle\cdot\rangle$ denotes an
ensemble average, we can obtain the frequency power spectra from the
time series measurements of magnetic fluctuations using the
following transform expression
\begin{equation}\label{eq3}
P_{ij}(f)=\int_{-\infty}^{\infty}R_{ij}^{(t)}(\tau)e^{-i2\pi
f\tau}d\tau.
\end{equation}
Substituting Equation \eqref{eq2} into Equation \eqref{eq3} and
using the Fourier transform, we can obtain
\begin{equation}\label{eq4}
P_{ij}(f)=\int
P_{ij}(\mathbf{k})\delta\left(f-\frac{\mathbf{k}\cdot\mathbf{V}}{2\pi}\right)d^3\mathbf{k},
\end{equation}
where the delta function $\delta$ can enforce the Doppler shift
condition between spatial and temporal frequencies. The feasibility
of this transform is rooted in the harmonic analysis of time series,
where each stationary time series can be depicted as a summation of
Fourier components \citep{Nieuwstadt2016}.

In the domain of wavevectors, the anisotropy is evident in the
structures of the energy spectrum, which encompass both parallel and
perpendicular components regarding the mean magnetic field
\citep{Narita2018}. Slab turbulence is characterized by wavevectors
exclusively aligned in the $\mathbf{B}_0$ direction, while 2D
turbulence is defined by wavevectors oriented purely orthogonal to
$\mathbf{B}_0$. According to the approach suggested by
\citet{Matthaeus1981}, the general spectral tensors can be divided
into two modes, i.e., the so-called slab model ($k_x=k_y=0$) and 2D
model ($k_{\parallel}=k_z=0$).

The slab model can be described as
\begin{equation}\label{eq5}
\begin{gathered}
P_{xx}^S(\mathbf{k})=P_{yy}^S(\mathbf{k})=G_S\left(k_z\right)\delta\left(k_x\right)\delta\left(k_y\right),\\
P_{zz}^S(\mathbf{k})=0.
\end{gathered}
\end{equation}
The 2D model can be expressed as
\begin{equation}\label{eq6}
\begin{gathered}
P_{xx}^{2D}(\mathbf{k})=\frac{G_{2D}\left(k_{\perp}\right)}{k_{\perp}^3}k_y^2\delta\left(k_z\right),\\
P_{yy}^{2D}(\mathbf{k})=\frac{G_{2D}\left(k_{\perp}\right)}{k_{\perp}^3}k_x^2\delta\left(k_z\right),\\
P_{zz}^{2D}(\mathbf{k})=0,
\end{gathered}
\end{equation}
where $k_{\perp}=\sqrt{k_{x}^{2}+k_{y}^{2}}$, $G_{S}$ is a function
of $k_{z}$, $G_{2D}$ is a function of $k_{\perp}$, and $\delta$
denotes the Dirac delta function. Substituting Equations \eqref{eq5}
and \eqref{eq6} into Equation \eqref{eq4}, we can shift the focus
from the wavevector domain to the frequency domain as demonstrated
by \citet{Bieber1996} and \citet{Zank2020}. The shifted slab model
can be written as
\begin{equation}\label{eq7}
\begin{gathered}
P_{xx}^S(f)=P_{yy}^S(f)=\frac{2\pi}{V\cos\psi}G_S\left(k_z\right),\\
k_z=\frac{2\pi f}{V\cos\psi}.
\end{gathered}
\end{equation}
The shifted 2D model can be described as
\begin{equation}\label{eq8}
\begin{gathered}
P_{xx}^{2D}(f)=\frac{4\pi}{V\sin\psi}\int_{\left|k_x\right|}^{\infty}\frac{G_{2D}\left(k_{\perp}\right)}{k_{\perp}^2}\sqrt{k_{\perp}^2-k_x^2}dk_{\perp},\\
P_{yy}^{2D}(f)=\frac{4\pi}{V\sin\psi}k_x^2\int_{\left|k_x\right|}^{\infty}\frac{G_{2D}\left(k_{\perp}\right)}{k_{\perp}^2}\frac{1}{\sqrt{k_{\perp}^2-k_x^2}}dk_{\perp},\\
k_x=\frac{2\pi f}{V\sin\psi}.
\end{gathered}
\end{equation}

Differentiating the Equation \eqref{eq8} with respect to the
frequency $f$, we can clarify the relationship between perpendicular
and parallel spectral components in the 2D model as
\begin{equation}\label{eq9}
P_{\perp}^{2D}(f)=-f\frac{dP_{\parallel}^{2D}(f)}{df}.
\end{equation}
In our coordinate system, the $y$-axis is orthogonal to the mean
magnetic field $\mathbf{B}_0$, which is aligned with the $z$-axis.
For the velocity vectors that lie within the $x$-$z$ plane, we can
project them onto the direction of the magnetic field. We designate
$P_{yy}(f)$ and $P_{xx}(f)$ as the perpendicular spectrum
$P_{\perp}(f)$ and the parallel spectrum $P_{\parallel}(f)$,
respectively. These components can serve as the principal diagonal
elements of the spectral tensor. Since $P_{zz}(f)=0$ for both
models, the total power spectrum can be simplified as
\begin{equation}\label{eq10}
P_{\text{total}}(f)=P^{S}(f)+P^{2D}(f)=P_{\parallel}(f)+P_{\perp}(f).
\end{equation}

It is usually assumed that $G_{S}$ in Equation \eqref{eq7} and the
integral term containing $G_{2D}$ in Equation \eqref{eq8} conform to
the power-law distributions
\citep[e.g.,][]{Bieber1996,Horbury2008,Forman2011,Zank2020}.
Specifically, we assume $G_S(k_z)=A_Sk_z^{-q}$, where $A_S$ is the
amplitude of slab turbulence, and $q$ is the spectral index. Then
from Equation \eqref{eq7} we can infer the following relation
\begin{equation}\label{eq11}
fP_{\perp}^S(f)=fP_{\parallel}^S(f)=A_Sk_z^{1-q}=A_S\left(\frac{2\pi
f}{V\cos\psi}\right)^{1-q}.
\end{equation}
For the 2D turbulence, we assume
$G_{2D}(k_\perp)=A_{2D}k_\perp^{-q}$, where $A_{2D}$ is the
amplitude of 2D turbulence. Evaluating the integrals in Equation
\eqref{eq8} and using Equation \eqref{eq9}, we can obtain
\begin{equation}\label{eq12}
\begin{aligned}
fP_{\parallel}^{2D}(f)&=A_{2D}\frac{2}{1+q}\left(\frac{2\pi f}{V\sin\psi}\right)^{1-q},\\
fP_{\perp}^{2D}(f)&=A_{2D}\frac{2q}{1+q}\left(\frac{2\pi
f}{V\sin\psi}\right)^{1-q}.
\end{aligned}
\end{equation}

For anisotropic solar wind turbulence based on the slab+2D composite
model, the individual spectra can be written as
\begin{equation}\label{eq13}
fP_{\perp}(f)=A_{S}\left(\frac{2\pi
f}{V\cos\psi}\right)^{1-q}+A_{2D}\frac{2q}{1+q}\left(\frac{2\pi
f}{V\sin\psi}\right)^{1-q},
\end{equation}
and
\begin{equation}\label{eq14}
fP_{\parallel}(f)=A_{S}\left(\frac{2\pi
f}{V\cos\psi}\right)^{1-q}+A_{2D}\frac{2}{1+q}\left(\frac{2\pi
f}{V\sin\psi}\right)^{1-q}.
\end{equation}
Dividing Equation \eqref{eq13} by Equation \eqref{eq14} yields
\begin{equation}\label{eq15}
\frac{P_{\perp}(f)}{P_{\parallel}(f)}=\frac{r(\cos\psi)^{q-1}+(1-r)\frac{2q}{1+q}(\sin\psi)^{q-1}}{r(\cos\psi)^{q-1}+(1-r)\frac{2}{1+q}(\sin\psi)^{q-1}},
\end{equation}
where $r$ denotes the slab fraction representing the energy in slab
mode relative to the total energy \citep{Bieber1996,Saur1999}, i.e.,
\begin{equation}\label{eq16}
r=\frac{A_{S}}{A_{S}+A_{2D}}.
\end{equation}

Combining Equations \eqref{eq13} and \eqref{eq14}, we can obtain the
total power spectrum as
\begin{equation}\label{eq17}
fP_{\text{total}}(f)=2A_{S}\left(\frac{2\pi
f}{V\cos\psi}\right)^{1-q}+2A_{2D}\left(\frac{2\pi
f}{V\sin\psi}\right)^{1-q}.
\end{equation}
On the right of Equation \eqref{eq17}, the first and second terms
represent the contributions from slab and 2D modes, respectively. To
examine the total energy spectrum at a specific reference frequency
$f_K$, we reformulate the above equation as
\begin{equation}\label{eq18}
f_KP_{\text{total}}(f_K)=E\left[r(\cos\psi)^{q-1}+(1-r)(\sin\psi)^{q-1}\right],
\end{equation}
where $E=2(A_{S}+A_{2D})k_{\text{ref}}^{1-q}$ denotes the total
amplitude at the reference wavenumber $k_{\text{ref}}$. In this
work, we pay attention to the measured frequencies ranging from
0.0026 Hz to 0.0547 Hz, with $f_K=0.0234$ Hz fixed, approximately
corresponding to $k_{\text{ref}}=3.7\times10^{-7}$ $m^{-1}$. We aim
to explore the power spectral anisotropy of solar wind turbulence
and its long-term (solar-cycle) variability through two independent
tests: the ratio test and the anisotropy test, as defined in
Equations \eqref{eq15} and \eqref{eq18}, respectively. Specifically,
we investigate the fraction of the slab component (as defined in
Equation \eqref{eq16}) in the solar wind turbulence and its
solar-cycle variations.

\section{Data and Methods} \label{sec:Data and Methods}
We utilize the magnetic field and solar wind data of 3-second
resolution collected by Wind spacecraft during the time period from
January 1995 to December 2023, which covers nearly three solar
cycles. The data of magnetic field and solar wind velocity were
provided by Magnetic Field Investigation \citep{Lepping1995} and
Three-Dimensional Plasma and Energetic Particle Investigation
\citep{Lin1995} instruments on board Wind spacecraft, respectively.
In this work, we do not aim to investigate the short-term features
of solar wind turbulence during some specific events, such as solar
energetic particle (SEP) events \citep{Bieber1996} and corotating
interaction regions \citep{Tessein2011}. Rather, we will investigate
the more general properties of the power spectral anisotropy of
solar wind turbulence in the context of solar activity and solar
cycle.

A critical aspect of our analyses is to choose an appropriate time
interval for determining the spectral properties of solar wind
turbulence. The time interval needs to be long enough to provide
reliable statistical estimates from sufficient observational data,
and meanwhile short enough to maintain the temporal correlation
required for effective correlation tensor calculations. Following
the operation in \citet{Wanner1993}, in the calculations we use
30-minute intervals, which are partly overlapping in our analyses.
Furthermore, each time interval should satisfy such requirements:
the missing data are less than $2\%$, and there are no consecutive
missing data points. We further use the linear interpolation method
to fill the small data gaps.

In the field-aligned coordinate system, we employ the pre-whitening
and post-darkening procedures to improve our spectral analysis
results \citep{Bieber1993}. The pre-whitening procedure is used to
compute the first-order difference between successive magnetic field
measurements for removing the mean field influence, and to isolate
the fluctuating components for subsequent analyses. The
autocorrelation function of the pre-whitened time series data is
then computed up to a time lag of 192 seconds. We further use the
discrete Fourier transform operation to obtain the power spectra of
the first-order difference of the magnetic fields. The
post-darkening corrections are then utilized to recover the original
spectral slopes, by minimizing the distortions caused by the
pre-whitening process.

We only retain the time intervals revealing a clear power-law
feature within the inertial range for further analyses. We discard
the power spectra containing negative spectral indices (i.e.,
$q<0$), which are false values arising from the noises during the
Blackman-Tukey analyses \citep{Blackman1958}. To avoid the influence
of potential aliasing, we discard the upper two thirds of the
frequencies in the power spectra. We pay attention to the remaining
21 frequency points from 0.0026 to 0.0547 Hz, corresponding to the
timescales approximately from 18 to 384 seconds.

In our analyses, the millions of time intervals in each year of
1995-2023 can provide us with sufficient samples for statistically
examining the 3D geometry of solar wind turbulence. We employ two
independent methods, i.e., the so-called spectrum ratio test and
spectrum anisotropy test, as described in Equations \eqref{eq15} and
\eqref{eq18}, respectively, to quantify the distribution of
turbulence energy in different wavevector directions. For the
spectrum ratio test, we calculate the ratio
$P_{\perp}/P_{\parallel}$ of the perpendicular to parallel power
spectra. Generally, it is assumed that the slab fraction $r$ is
independent of the frequencies in the frequency range of interest
\citep{Bieber1996,Leamon1998}. The ratio $P_{\perp}/P_{\parallel}$
is computed at a fixed frequency of 0.0234 Hz, which is the midpoint
of the considered frequency range. A ratio value close to unity
implies that the turbulence mode with wavevectors parallel to the
mean magnetic field is dominant, whereas the higher values suggest
larger contributions from perpendicular wavevectors. We employ the
anisotropy test to evaluate the dependence of the normalized total
energy spectrum $[f_KP_{\text{total}}(f_K)]_{\text{NORM}}$ on the
field-to-flow angle $\psi$. \citet{Bieber1996} assumed that the
normalized total energy $E$ is independent of the angle $\psi$.
However, the observational evidence presented in \citet{Saur1999}
revealed that $E$ roughly doubles when the angle $\psi$ increases
from $0^\circ$ to $90^\circ$. Therefore, in this work we regard $E$
as dependent on the angle $\psi$. For both tests, the field-to-flow
angle $\psi$ ranging from $0^\circ$ to $90^\circ$ is divided into
nine bins with equal width ($10^\circ$). Within each bin, we
calculate the geometric means of the spectral ratios and normalize
the total energy spectrum values to characterize the geometry of the
solar wind turbulence.

\section{Results and Discussion} \label{sec:Results and Discussion}
\subsection{Geometry of the Solar Wind Turbulence}
We analyze the geometry of the solar wind turbulence during a time
period of nearly three solar cycles (1995-2023). For each year, we
calculate the geometric mean of the ratio $P_{\perp}/P_{\parallel}$
of the perpendicular to parallel power spectral densities at the
frequency of 0.0234 Hz. Table \ref{tab:t1} shows the ratio values in
each year, which range from 1.38 to 1.59, with an overall geometric
mean of 1.51. Our results agree well with the ratio value $\sim1.40$
reported by \citet{Bieber1996}, reinforcing the thought that the
solar wind turbulence should be preferably described by the slab+2D
two-component composite model. In the slab+2D composite model, the
pure slab mode ($r=1$) corresponds to the ratio value
($P_{\perp}/P_{\parallel}$) of unity, while the pure 2D mode ($r=0$)
corresponds to the specific ratio value equal to the spectral index
$q$ ($q=5/3$, if a Kolmogorov-like spectrum is assumed). These
relations can be easily inferred in Equation \eqref{eq15}. The
values of ratio $P_{\perp}/P_{\parallel}$ shown in Table
\ref{tab:t1} indicate that the 2D component is approximately
dominant in the solar wind turbulence.

\begin{figure}
 \epsscale{1.0}
 \plotone{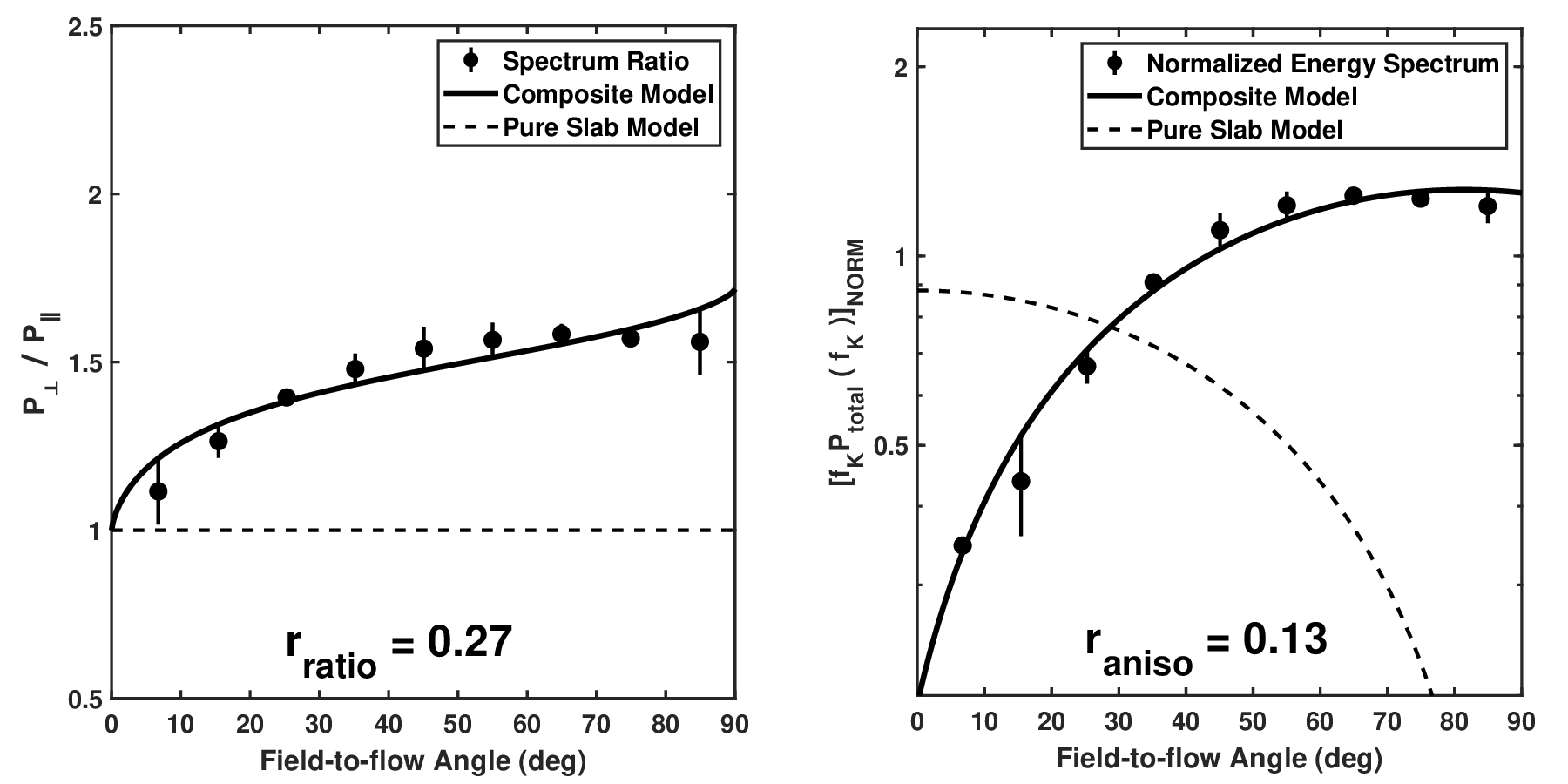}
 \caption{Left: ratio $P_{\perp}/P_{\parallel}$ of perpendicular to parallel power spectra
 as a function of field-to-flow angle $\psi$. The composite model (solid line) consisting of $73\%$ 2D and $27\%$ slab
 turbulence is consistent with the observations (data points). The dashed line
 denotes the pure slab model ($r=1$), which significantly deviates from the
 observations. Right: normalized total energy spectrum $\left[f_{K}P_{\text{total}}(f_{K})\right]_{\text{NORM}}$ as a function of field-to-flow angle
 $\psi$. The composite model (solid line) consisting of $87\%$ 2D and $13\%$ slab turbulence
 is consistent with the observations (data points). Again, the pure slab model ($r=1$, dashed line) significantly deviates from the realistic
 observations. \label{fig:f1}}
\end{figure}

Figure \ref{fig:f1} presents the results of the spectrum ratio test
and spectrum anisotropy test using the data during the period
1995-2023. The left panel of Figure \ref{fig:f1} shows the ratio
$P_{\perp}/P_{\parallel}$ as a function of the field-to-flow angle
$\psi$. As we can see, the ratio initially increases with increasing
$\psi$ and peaks between $60^\circ$ and $70^\circ$. The solid line
denotes the fitting result of the slab+2D composite model, which
determines the value of slab fraction, i.e.,
$r_{\text{ratio}}=0.27$. This indicates that approximately $73\%$ of
the turbulence energy is associated with the 2D component, and the
remaining $27\%$ of the energy is associated with the slab
component. Therefore, the solar wind turbulence is 2D-component
dominated, which agrees with the previous studies. The dashed line
denotes the prediction of pure slab model (i.e., $r\equiv1$), which
systematically underestimates the observational values of the ratio
$P_{\perp}/P_{\parallel}$ for all the field-to-flow angles $\psi$.
This discrepancy reveals the inadequacy of the pure slab model for
describing the realistic geometry of the solar wind turbulence.
Instead, the two-component composite model dominated by the 2D mode
is much more appropriate for depicting the observations.

The right panel of Figure \ref{fig:f1} presents the normalized total
energy spectrum $[f_KP_{\text{total}}(f_K)]_{\text{NORM}}$ as a
function of the field-to-flow angle $\psi$. The normalized
turbulence power increases by a factor of $\sim3$ as $\psi$ varies
from $0^\circ$ to $90^\circ$. Fitting the composite model to the
observations gives a slab fraction $r_{\text{aniso}}=0.13$, which
means that roughly $87\%$ of the turbulence energy is associated
with the 2D component. As in the left panel, the pure slab model
($r=1$, dashed line) almost completely deviates from the
observations. Therefore, the slab+2D composite model is more
reasonable for interpreting the observations. As we can see, both
the tests reveal the similar small slab fraction values, which
indicates that the 2D component is dominant in the solar wind
turbulence. Our results are in agreement with the previous studies
\citep{Bieber1996,Leamon1998}.

\subsection{Solar-cycle Variability of Composite Geometry}
We analyze the long-term variation of the slab fraction during the
29-year time period to explore the solar-cycle variability of the
composite geometry in the solar wind turbulence. In addition to
ratio $P_{\perp}/P_{\parallel}$, Table \ref{tab:t1} also presents
other key quantities for each year during 1995-2023: mean
field-to-flow angle $\psi$, slab fraction $r_{\text{ratio}}$ derived
from the spectrum ratio test, slab fraction $r_{\text{aniso}}$
derived from the spectrum anisotropy test, and the averaged slab
fraction $\bar{r}$ of $r_{\text{ratio}}$ and $r_{\text{aniso}}$. The
slab fraction $r_{\text{ratio}}$ ranges from 0.17 to 0.39, and
$r_{\text{aniso}}$ ranges from approximately 0.00 to 0.24. We can
see that the slab fraction values from the spectrum ratio test are
systematically larger than those from the spectrum anisotropy test.
The most pronounced discrepancy occurs in 1996, where
$r_{\text{aniso}}=0$ and $r_{\text{ratio}}=0.28$. The overall value
of $r_{\text{ratio}}$ (0.27) is approximately twice that of
$r_{\text{aniso}}$ (0.13) for the time period 1995-2023. The
averaged slab fraction $\bar{r}$ of $r_{\text{ratio}}$ and
$r_{\text{aniso}}$ suggests that approximately $71\%-87\%$ of the
total turbulence energy within the inertial range is distributed in
the 2D component. That is to say, for quite a long time period
(1995-2023), the 2D component almost always remains dominant in the
energy anisotropy of the solar wind turbulence.

\begin{figure}
 \epsscale{1.0}
 \plotone{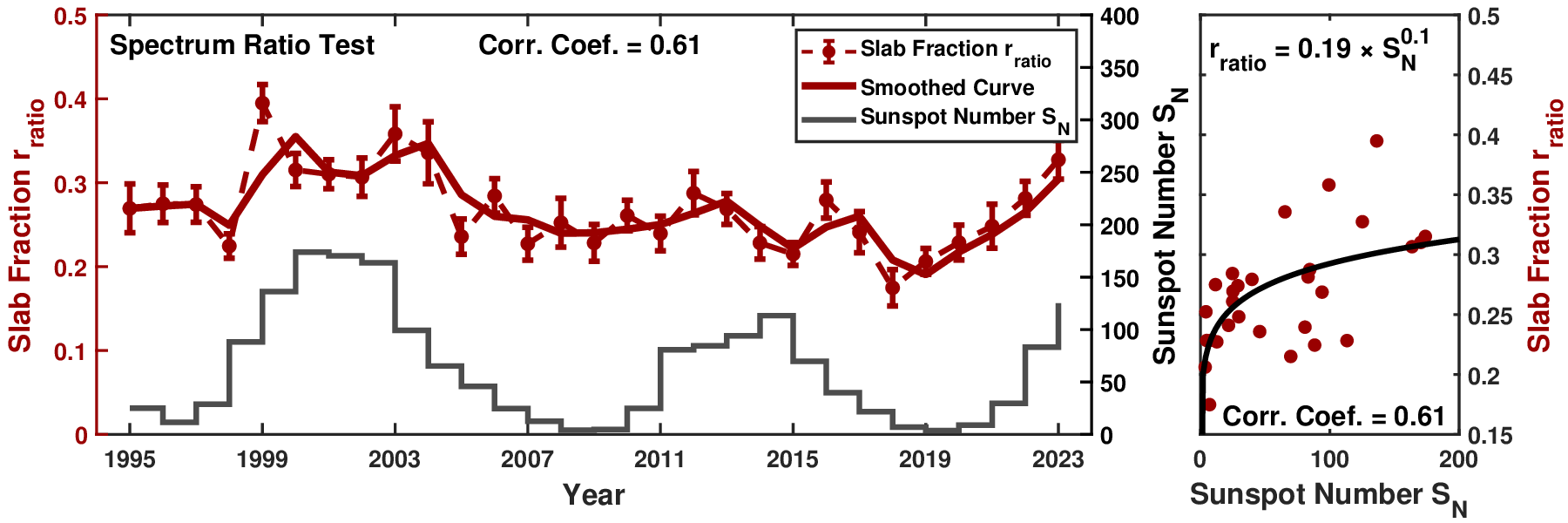}
 \caption{Left: temporal evolutions of the sunspot number $S_{N}$ (grey line) and the slab fraction $r_{\text{ratio}}$ (red circles)
 derived from the spectrum ratio test during the time period
 1995-2023. The red solid line denotes the smoothed results of
 $r_{\text{ratio}}$. Right: variations of the slab fraction $r_{\text{ratio}}$ with the increase
of the sunspot number. The correlation coefficient between
$r_{\text{ratio}}$ and the sunspot number is 0.61. \label{fig:f2}}
\end{figure}

The left panel of Figure \ref{fig:f2} shows the temporal evolutions
of the slab fraction $r_{\text{ratio}}$ derived from the spectrum
ratio test and the sunspot number during the period 1995-2023. The
slab fraction $r_{\text{ratio}}$ generally fluctuates around 0.27,
with a minimum of 0.17 in 2018. Interestingly, the larger slab
fraction values tend to coincide with the increased sunspot numbers,
and the smaller slab fraction values roughly accompany the decreased
sunspot numbers. This phenomenon indicates that the solar activity
influences the geometry of the solar wind turbulence. The right
panel of Figure \ref{fig:f2} shows the variations of slab fraction
$r_{\text{ratio}}$ with the increase of the sunspot number. The
correlation coefficient between $r_{\text{ratio}}$ and the sunspot
number is 0.61. This finding means that the intense solar activity
generally enhances the field-aligned turbulence. However, the 2D
component always remains dominant, even during the periods of solar
maxima. The empirical functional relation (black curve) between the
slab fraction $r_{\text{ratio}}$ and the sunspot number $S_{N}$ can
be described as
\begin{equation}\label{eq19}
r_{\text{ratio}}=0.19\times S_{N}^{0.1}.
\end{equation}

\begin{figure}
 \epsscale{1.0}
 \plotone{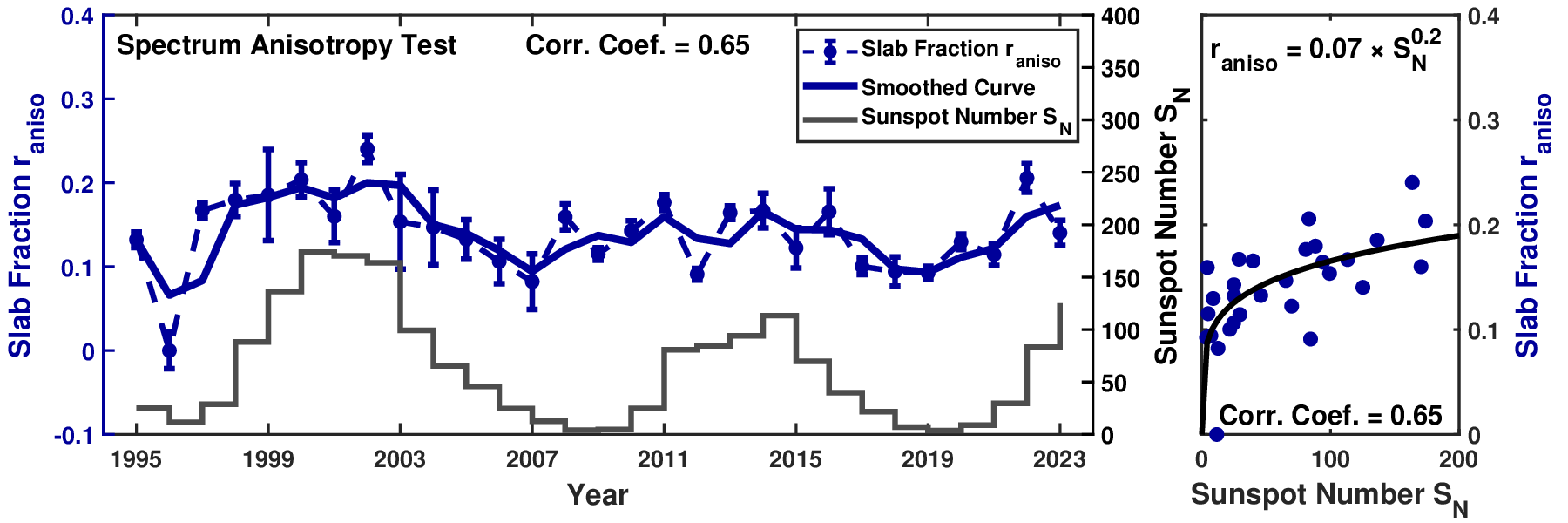}
 \caption{Left: evolutions of the sunspot number $S_{N}$ (grey line) and the slab fraction
 $r_{\text{aniso}}$ (blue circles) derived from the spectrum anisotropy test during
 the period 1995-2023. The blue solid line denotes the smoothed results
 of $r_{\text{aniso}}$. Right: variations of the slab fraction
 $r_{\text{aniso}}$ with the increase of the sunspot number. The correlation coefficient
 between $r_{\text{aniso}}$ and the sunspot number is 0.65. \label{fig:f3}}
\end{figure}

The left panel of Figure \ref{fig:f3} presents the temporal
evolutions of the slab fraction $r_{\text{aniso}}$ derived from the
spectrum anisotropy test and the sunspot number during 1995-2023.
Generally, the slab fraction $r_{\text{aniso}}$ fluctuates around
0.15, which suggests that roughly $85\%$ of the turbulence energy in
the inertial range is associated with the wavevectors perpendicular
to the mean magnetic field. The right panel of Figure \ref{fig:f3}
displays the variations of slab fraction $r_{\text{aniso}}$ with the
increase of the sunspot number. The correlation coefficient between
$r_{\text{aniso}}$ and the sunspot number is 0.65. As mentioned
above, the intense solar activity enhances the turbulence of the
field-aligned component. Nevertheless, the essential manifestation
is still that the 2D component remains dominant in the solar wind
turbulence throughout the entire solar cycles. The empirical
relation (black curve) between the slab fraction $r_{\text{aniso}}$
and the sunspot number $S_{N}$ can be expressed as
\begin{equation}\label{eq20}
r_{\text{aniso}}=0.07\times S_{N}^{0.2}.
\end{equation}

The above consistent results from the two types of test methods show
that the solar activity accompanying the solar cycles significantly
influences the 3D geometry of solar wind turbulence in the inertial
range. Specifically, the slab component in the solar wind turbulence
is enhanced during the rise phases of the solar cycles. The increase
of slab fraction is likely due to the intensified interplanetary
magnetic field magnitude during the rise phases of the solar
activity cycles. As we know, the Alfv{\'e}n speed $v_{A}$ is related
to the interplanetary magnetic field magnitude $B_0$ as
\begin{equation}\label{eq21}
v_{A}=\frac{B_0}{\sqrt{\mu_0\varrho_0}},
\end{equation}
where $\mu_0$ is the permeability, and $\varrho_0$ is the plasma
density. Thus, the Alfv{\'e}n speed $v_{A}$ increases with the
increasing magnetic field magnitude $B_0$ during the rise phases of
the solar cycles. Generally, the larger Alfv{\'e}n speed indicates
the larger slab fraction in the solar wind turbulence. Therefore,
the slab fraction increases with the increasing sunspot number
during the rise phases of the solar cycles.

\section{Summary}\label{sec:Summary}
In this work, we investigate the temporal evolutions of the 3D
geometry of solar wind turbulence and particularly its solar-cycle
variability during the time period 1995-2023. We use the so-called
spectrum ratio test and spectrum anisotropy test to analyze the
magnetic field and solar wind data of 3-second resolution collected
by Wind spacecraft. The distribution of turbulence energy between
the slab and 2D components in the inertial range is determined. Our
results show that the solar wind turbulence is 2D-component
dominated, i.e., $\sim80\%$ of the turbulence energy is in the 2D
component. Our results agree well with the previous studies
\citep{Bieber1996,Leamon1998}, which reinforces the viewpoint that
most turbulence energy is associated with the wavevectors
perpendicular to the mean magnetic field.

More significantly, both the ratio test and anisotropy test reveal
that there exists a positive correlation between the slab fractions
and the sunspot numbers accompanying the solar activity cycles. The
correlation coefficients between the slab fractions and the sunspot
numbers determined with the ratio test and anisotropy test are 0.61
and 0.65, respectively. That is to say, the slab fraction is
enhanced during the rise phases (with increasing sunspot number) of
the solar cycles. These findings show that the solar activity
accompanying the solar cycles considerably influences the slab
fraction (also 2D fraction) and the 3D composite geometry of the
solar wind turbulence. The enhancement of slab fraction is
associated with the intensified interplanetary magnetic field
magnitude and consequently the increased Alfv{\'e}n speed $v_{A}$
during the rise phases of the solar cycles.

Our results presented in this work will be valuable for achieving an
in-depth understanding of the 3D composite geometry and spectral
anisotropy of the solar wind turbulence, and particularly of their
solar-cycle variability. We suggest that the modulation effects of
solar activity and solar cycles should be taken into account when we
refine the solar wind turbulence models and especially the
anisotropy models. The knowledge of solar wind turbulence and its
anisotropy is crucial for the studies of several important topics in
space physics and heliophysics, such as cosmic ray propagation and
modulation, SEP acceleration and transport, and solar wind heating
and acceleration, etc. Therefore, our findings will provide some new
insights into these traditional subjects. In the future, we shall
further investigate the solar wind turbulence anisotropy and its
solar-cycle variations, so as to lay physical foundations for this
interesting subject and the relevant topics.


\acknowledgments

This work was supported in part by the B-type Strategic Priority
Program of the Chinese Academy of Sciences under grant XDB41000000,
and the National Natural Science Foundation of China under grant
41874207. H.-Q. He gratefully acknowledges the partial support of
the Youth Innovation Promotion Association of the Chinese Academy of
Sciences (No. 2017091). We benefited from the data of Wind provided
by NASA/Space Physics Data Facility (SPDF)/CDAWeb. The sunspot data
were provided by the World Data Center SILSO, Royal Observatory of
Belgium, Brussels.


\clearpage



\begin{table}
    \caption{The computation results of the key quantities during the time
    period 1995-2023: mean field-to-flow angle $\psi$, geometric mean of
    $P_{\perp}/P_{\parallel}$, slab fraction $r_{\text{ratio}}$ derived from the spectrum ratio test,
    slab fraction $r_{\text{aniso}}$ derived from the spectrum anisotropy test, and the averaged slab fraction $\bar{r}$ of $r_{\text{ratio}}$ and $r_{\text{aniso}}$. \label{tab:t1}}
    \centering
    \newsavebox{\tablebox}
    \begin{lrbox}{\tablebox}
    \begin{tabular}{cccccc}
\hline\hline
Year & $\psi$ ($^\circ$) & $P_{\perp}/P_{\parallel}$ & $r_{\text{ratio}}$ & $r_{\text{aniso}}$ & $\bar{r}$ \\
\hline
1995-2023 & 52.86 & 1.51 & 0.27 & 0.13 & 0.20 \\
\hline
1995 & 51.75 & 1.49 & 0.27 & 0.13 & 0.20 \\
1996 & 48.66 & 1.43 & 0.28 & 0.00 & 0.14 \\
1997 & 53.20 & 1.50 & 0.27 & 0.17 & 0.22 \\
1998 & 54.17 & 1.53 & 0.22 & 0.18 & 0.20 \\
1999 & 53.08 & 1.41 & 0.39 & 0.19 & 0.29 \\
2000 & 54.03 & 1.48 & 0.32 & 0.20 & 0.26 \\
2001 & 54.63 & 1.47 & 0.31 & 0.16 & 0.24 \\
2002 & 53.89 & 1.49 & 0.31 & 0.24 & 0.27 \\
2003 & 48.77 & 1.38 & 0.36 & 0.15 & 0.26 \\
2004 & 49.77 & 1.42 & 0.34 & 0.15 & 0.24 \\
2005 & 51.68 & 1.52 & 0.24 & 0.13 & 0.18 \\
2006 & 53.71 & 1.51 & 0.28 & 0.11 & 0.20 \\
2007 & 51.41 & 1.50 & 0.23 & 0.08 & 0.15 \\
2008 & 52.10 & 1.50 & 0.25 & 0.16 & 0.21 \\
2009 & 54.91 & 1.58 & 0.23 & 0.12 & 0.17 \\
2010 & 53.99 & 1.55 & 0.26 & 0.14 & 0.20 \\
2011 & 53.48 & 1.52 & 0.24 & 0.18 & 0.21 \\
2012 & 56.91 & 1.59 & 0.29 & 0.09 & 0.19 \\
2013 & 55.25 & 1.59 & 0.27 & 0.16 & 0.22 \\
2014 & 55.65 & 1.59 & 0.23 & 0.17 & 0.20 \\
2015 & 52.66 & 1.55 & 0.22 & 0.12 & 0.17 \\
2016 & 52.18 & 1.49 & 0.28 & 0.17 & 0.22 \\
2017 & 50.26 & 1.48 & 0.24 & 0.10 & 0.17 \\
2018 & 51.11 & 1.51 & 0.17 & 0.09 & 0.13 \\
2019 & 51.83 & 1.50 & 0.21 & 0.09 & 0.15 \\
2020 & 52.10 & 1.51 & 0.23 & 0.13 & 0.18 \\
2021 & 54.46 & 1.55 & 0.25 & 0.11 & 0.18 \\
2022 & 54.11 & 1.53 & 0.28 & 0.21 & 0.24 \\
2023 & 55.35 & 1.54 & 0.33 & 0.14 & 0.23 \\
\hline
\end{tabular}
\end{lrbox}
\resizebox{0.4\textwidth}{!}{\usebox{\tablebox}}
\end{table}
\clearpage


\clearpage

\clearpage

\clearpage


\end{document}